\documentclass{article}

\usepackage{arxiv}

\usepackage[utf8]{inputenc} 
\usepackage[T1]{fontenc}    
\usepackage{hyperref}       
\usepackage{url}            
\usepackage{booktabs}       
\usepackage{amsfonts}       
\usepackage{amsmath}        
\usepackage{nicefrac}       
\usepackage{microtype}      
\usepackage{textgreek}      
\usepackage{listings}
\usepackage{graphicx}
\usepackage{natbib}
\usepackage{doi}
\usepackage{tikz}
\newtheorem{theorem}{Theorem}

\title{A Wave-Geometric Duality for Hyperdimensional Computing \\ 
       \large Unitary embedding, primitive-wave realization, and electromagnetic validation }

\author{ \href{https://orcid.org/0009-0005-1235-6295}{\includegraphics[scale=0.06]{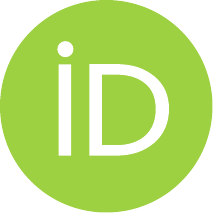}\hspace{1mm}Tyler L. Poore} \\
	Independent Researcher \\
	Sunnyvale, CA 94083 \\
}

\renewcommand{\shorttitle}{Wave-Geometric Duality for HDC}

\hypersetup{
pdftitle={A Wave-Geometric Duality for Hyperdimensional Computing: Exact Embedding, Primitive-Wave Realization, and Electromagnetic Validation},
pdfsubject={cs.ET (Emerging Technologies); hyperdimensional computing; unitary wave embedding; primitive-wave duality},
pdfauthor={Tyler L. Poore},
pdfkeywords={cs.ET, emerging technologies, hyperdimensional computing, vector symbolic architectures, unitary embedding, bundling, binding, permutation, power-similarity bridge, electromagnetic validation, resonant field computing},
}

\begin{document}
\maketitle

\begin{abstract}
    Hyperdimensional computing (HDC), also referred to as vector symbolic architectures (VSA), represents information with high-dimensional vectors and a compact algebra of primitives. This paper establishes an explicitly unitary embedding from discrete bipolar HDC/VSA vectors to coherent broadband waveforms and develops a common wave-domain realization of the core HDC/VSA primitives within that embedding. Under the resulting RFC/UWE stack, bundling becomes linear superposition, permutation becomes coherent phase evolution, binding is reproduced by nonlinear spectral mixing together with an engineered aliasing step that restores circular-convolution structure, and similarity is recovered as a calibrated differential-power readout. The similarity/readout layer receives the deepest treatment because it determines how the stack becomes physically measurable and what the architecture must satisfy, not because it stands apart from the rest of the framework. Full-wave FDTD studies validate the physically nontrivial parts of this program, including the array-level readout mechanism and the binding pipeline under realistic propagation. In a documented $N=1000$ mutually coupled-array calibration, the predicted interaction effect appears with the expected sign pattern and order of magnitude, yielding a coupled Correlation Contrast Ratio of approximately $\mathrm{CCR}_{\mathrm{cpl}} \approx 8.7\times 10^{-5}$. A controlled anisotropic-isolation surrogate further shows that reducing coupling-dominated baseline power lifts contrast dramatically, clarifying why emitter isolation is an architectural requirement rather than an optional refinement. The result is a wave--geometric duality for HDC/VSA: existing symbolic operations admit a physically grounded waveform realization, while coherence, isolation, and readout sensitivity remain the central engineering constraints for future hardware.
\end{abstract}

\section{Introduction}

Hyperdimensional computing (HDC), also known as vector symbolic architectures (VSA), represents information with high-dimensional random vectors and a small algebra of primitives, and it has been valued for robustness, compositionality, and statistical predictability in high dimensions \cite{kanerva2009}. Throughout, I use \emph{HDC/VSA} as an umbrella term for this literature. Most work treats hypervectors as abstract algebraic objects, with implementations relying on von Neumann architectures where throughput is limited by memory movement and the per-operation cost that scales with dimensionality \cite{kleyko2022hardware}.

Hardware accelerators have been developed to mitigate these limitations. Photonic systems such as PhotoHDC use electro-photonic hardware to accelerate HDC training and inference across optical channels \cite{fayza2023photonics}. These designs achieve high throughput, but they still instantiate an externally specified algebra rather than showing that the algebra follows directly from physical principles.

Most published HDC hardware to date fits this accelerator/emulation pattern: the symbolic operations are specified first and then mapped onto the substrate. On the stricter criterion that the algebra itself be shown to be native to the physics rather than merely executed efficiently by it, the literature remains comparatively sparse \cite{kleyko2022hardware}.

This work develops a framework that unifies the algebra of HDC with wave mechanics. A unitary mapping is constructed that sends discrete bipolar hypervectors to continuous broadband waveforms while preserving inner products. Under this mapping, a vector's information is encoded in the coherent structure of its spectral components. Bundling corresponds to linear superposition, binding is reproduced by nonlinear spectral mixing combined with an engineered aliasing step that matches circular convolution, and permutation becomes coherent phase evolution. The mapping is norm-preserving, invertible under stated conditions, and consistent with the geometric structure of the discrete hyperspace.

A particularly consequential result within this framework is a direct relationship between vector similarity and a physically measurable quantity. Under the unitary embedding, the differential power produced by the interaction of two waveforms is linearly proportional to the cosine similarity of the corresponding hypervectors. This establishes a physical mechanism for similarity detection via a single quadratic power measurement. The required conditions on coherence, spectral structure, and invertibility are derived explicitly.

\paragraph{Primary claim and scope.}
The primary claim of this manuscript is simple: there exists an explicitly unitary embedding from bipolar HDC/VSA vectors to coherent waveforms under which the core primitives admit a common wave-domain realization---bundling as superposition, permutation as coherent phase evolution, binding via nonlinear spectral mixing plus engineered aliasing, and similarity via calibrated differential power. The similarity/readout layer receives the deepest treatment in this draft because it determines how the stack becomes physically measurable and what the architecture must satisfy, not because it displaces the rest of the RFC/UWE framework. To avoid scope ambiguity, this manuscript separates three layers: (i) HDC/VSA as the symbolic algebra (superposition, binding, permutation, similarity), (ii) UWE as the unitary embedding that maps vectors to waveforms while preserving geometry, and (iii) RFC as the physical architecture and readout protocol that executes those operations in wave hardware via measurable power interactions. Under this definition, RFC is not introduced as a replacement symbolic algebra; it is a substrate-level instantiation of existing HDC/VSA algebra (with MAP-B-style bipolar operations as the baseline in this draft). Because UWE is invertible and norm-preserving, the same computation can be viewed equivalently in discrete vector space or continuous wave space---a dual representation rather than two unrelated formalisms. More precisely, the manuscript argues for a candidate algebra-revealing instantiation under an explicit embedding and readout protocol; it does not claim that raw electromagnetics, absent that embedding and the required aliasing/projection steps, automatically realizes every HDC primitive. Sections on architecture, emitter isolation, and hardware are included to delimit what a physical realization would need, not to claim a complete device stack.

During development, many adjacent parameter choices and simulation variants were explored. The exact quantitative values quoted in this manuscript are tied only to the representative documented configurations stated in the methodology and appendices below; the broader parameter sweeps motivate the engineering discussion but are not used as hidden provenance for the headline numbers.

Full-wave three-dimensional FDTD simulations validate the physically nontrivial parts of the framework emphasized in this draft. At the array-readout level, these simulations confirm the expected sign pattern and magnitude of the interaction effect, yielding a Correlation Contrast Ratio (CCR) of approximately \(8.7 \times 10^{-5}\) for \(N = 1000\). The effect persists under various physical configurations and is consistent with the analytical model.

\textbf{Contributions.} The contributions of this work are as follows:

\begin{itemize}
    \item An explicitly unitary embedding (UWE) from bipolar hypervectors to coherent broadband waveforms that preserves norms and inner products.
    \item An RFC/UWE primitive map under which bundling, permutation, binding, and similarity acquire explicit wave-domain realizations within a common substrate.
    \item A Power--Similarity Bridge showing that differential power is linear in vector inner product, and therefore in cosine similarity for normalized vectors, together with the resulting readout constraints on coherence, isolation, and measurement sensitivity.
    \item Electromagnetic validation of the framework's physically nontrivial mechanisms, including array-level readout in a mutually coupled setting and physically propagated binding under the stated aliasing plan.
\end{itemize}

Accordingly, the manuscript should be read primarily as a theory-and-validation paper; implementation discussions are included to state the boundary conditions for future realization.

All results assume mutually phase-locked tone combs; a quantified phase-jitter tolerance study is provided in Section~\ref{sec:jitter-binding} (see also Appendix~\ref{app:jitter}).

\section{Related Work}
\label{sec:related}

Hyperdimensional computing (HDC), also known as Vector Symbolic Architectures (VSA), has historically developed along two largely independent axes: (i) abstract algebraic models of high-dimensional representation and (ii) hardware efforts that accelerate or instantiate these models. The present framework is situated relative to both lineages but departs from each in a specific way: rather than merely emulating the algebra on a physical substrate, it asks whether an explicit embedding can make core HDC operations readable \emph{from} wave physics itself. This section positions the present contribution within the broader HDC/VSA literature and within prior physical and wave-based implementations.

\subsection{Vector Symbolic Architectures and Hyperdimensional Computing}

Kanerva’s Binary Spatter Code line of work~\cite{kanerva2009} established the statistical geometry of high-dimensional random representations, where quasi-orthogonality and concentration of measure enable robust associative memory. In practice, superposition in this family is implemented by component-wise accumulation followed by majority/sign decoding, and binding is implemented by XOR in binary form (or equivalently by component-wise multiplication in bipolar form). The key point for this paper is that these operators are defined algebraically and are usually realized on digital hardware, independent of any specific physical wave mechanism.

Plate’s Holographic Reduced Representations (HRR)~\cite{plate1995} are a real-valued VSA formulation in which binding is circular convolution. By the convolution theorem, circular convolution in vector space corresponds to pointwise multiplication in the Fourier domain. This mathematical correspondence is closely related to the present work. The distinction is implementation: HRR typically computes convolution algorithmically (e.g., FFT pipelines), whereas this work uses the Unitary Wave Embedding (UWE) to show how binding, superposition, and permutation can be instantiated by wave-domain processes (spectral mixing, linear superposition, and coherent phase evolution).

Gayler’s Multiply--Add--Permute (MAP) framework~\cite{gayler2004} abstracted away from specific data types and identified the structural requirements any substrate must satisfy: a superposition operator that preserves similarity, a binding operator that produces dissimilar composites, and a permutation operator to encode role or position. In this manuscript, UWE denotes the embedding map itself, while RFC denotes the full physical computing architecture built on that map. Under this distinction, RFC satisfies MAP-style requirements by using Maxwell-compatible wavefields rather than constructing digital operators.

Sparse distributed representations~\cite{rachkovskij2001} emphasize biological plausibility and memory efficiency. The focus here is different: for wave-domain readout, dense comb occupancy can improve coherent integration and effective SNR under fixed bandwidth, making dense encodings a practical choice for this RFC formulation. Accordingly, UWE is positioned as a physical instantiation pathway for dense VSA geometry rather than a new sparse representational family.

Dynamic VSA approaches such as Resonator Networks~\cite{frady2019r} implement iterative attractor dynamics to factor or retrieve bound structures. Although the RFC framework developed here does not rely on dynamical relaxation, its key mechanism, energy exchange and interference between coherent fields, is closely related to the physical processes that could underlie resonator-like computation.

Kleyko et al.~\cite{kleyko2022survey} provide a comprehensive taxonomy of HDC/VSA models and data transformations. In that context, the present contribution is best viewed as a physically grounded instantiation of Fourier-domain VSA operations: the symbolic algebra remains familiar, while the novelty lies in the explicit invertible mapping to wavefields and the direct physical realization of those operations. In this sense, RFC is not introduced here as a replacement symbolic algebra; rather, it is a substrate-level realization layer (demonstrated for MAP-B style bipolar operations in this manuscript) that could host other compatible VSA algebras under the same unitary embedding principles.

\subsection{Physical, Photonic, and RF Implementations of HDC}

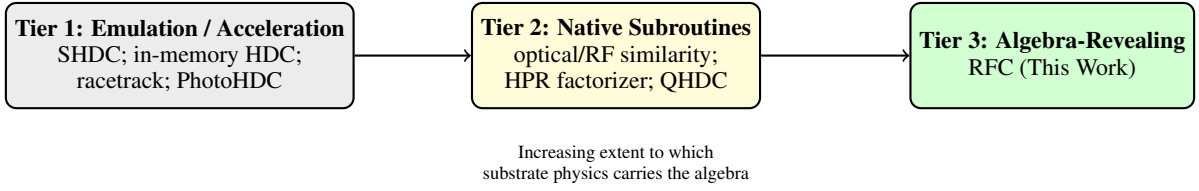
\begin{figure}[t]
\centering

\begin{tikzpicture}[
    node/.style={
        rounded corners,
        draw,
        thick,
        align=center,
        font=\small,
        minimum width=3.7cm,
        minimum height=1.4cm
    }
]

\node[node, fill=gray!15] (tier1) at (0,0) {
    \textbf{Tier~1: Emulation / Acceleration}\\
    \footnotesize SHDC; in-memory HDC;\\
    \footnotesize racetrack; PhotoHDC
};

\node[node, fill=yellow!18] (tier2) at (5.8,0) {
    \textbf{Tier~2: Native Subroutines}\\
    \footnotesize optical/RF similarity;\\
    \footnotesize HPR factorizer; QHDC
};

\node[node, fill=green!18] (tier3) at (11.6,0) {
    \textbf{Tier~3: Algebra-Revealing}\\
    \footnotesize RFC (This Work)
};

\draw[->, thick] (tier1) -- (tier2);
\draw[->, thick] (tier2) -- (tier3);

\node[align=center, font=\scriptsize] at (5.8,-1.45) {
    Increasing extent to which\\
    substrate physics carries the algebra
};

\end{tikzpicture}

\caption{
One useful way to organize hyperdimensional computing hardware approaches by where the algebra is realized, rather than by device family alone.
Tier~1 executes pre-specified HDC operators efficiently in hardware.
Tier~2 aligns specific HDC-relevant operations with native substrate physics, but does not yet establish a full intrinsic algebra.
Tier~3, including the proposed Resonant Field Computing (RFC), argues for an explicit embedding under which multiple HDC primitives are carried by the medium itself.
The taxonomy is organizational rather than chronological.
}
\label{fig:hdc_taxonomy}
\end{figure}

Recent HDC hardware papers are better organized by \emph{where the algebra lives} than by device family alone. Following the abstraction-layer perspective of Kleyko et al.~\cite{kleyko2022hardware}, I distinguish three cases relevant to the present manuscript. In \emph{algebra-preserving emulation/acceleration}, the HDC/VSA primitives are defined symbolically and then executed more efficiently in hardware, often still through component-wise or channel-wise dataflow. In \emph{physics-native subroutines / partial primitive alignment}, a native physical effect directly realizes one or more HDC-relevant operations, but the work does not establish that the full algebra is intrinsic to the medium. In \emph{algebra-revealing instantiation}, an explicit embedding or equivalence shows that multiple HDC primitives arise from the native dynamics of the substrate itself. Figure~\ref{fig:hdc_taxonomy} summarizes this organization and situates RFC in the third category. The figure is organizational rather than chronological.

\paragraph{Algebra-preserving emulation and in-memory acceleration.}
Superconducting HDC (SHDC) pushes emulation to its thermodynamic limits by using Rapid Single-Flux-Quantum logic to implement hyperdimensional associative-memory circuits~\cite{huch2023shdc}. In MAP-B-like settings, binding remains a component-wise XOR realized by RSFQ logic gates, so hardware complexity still scales with dimensionality. The architecture achieves extremely high clock rates and low switching energy, but requires cryogenic cooling and retains the $O(N)$ scaling inherent to component-wise emulation.

In-memory HDC with phase-change memory arrays implements a complete HDC system using memristive crossbar engines together with peripheral CMOS circuitry~\cite{karunaratne2020imemory}. Ferroelectric in-memory HDC similarly uses FeFET-based multi-bit IMC/CAM structures to reproduce HDC encoding and associative search with improved efficiency~\cite{kazemi2022fefet}. Racetrack-memory HDC conducts the framework within memory by using transverse-read operations to realize XOR and addition with minimal extra CMOS~\cite{khan2022racetrack}.

Photonic HDC platforms, such as PhotoHDC~\cite{fayza2023photonics}, use electro-photonic hardware to accelerate HDC training and inference and support multiple HDC encoding schemes. Across these works, the hardware contribution is substantial, but the algebra remains externally specified: the substrate accelerates pre-defined HDC operators rather than experimentally demonstrating that the HDC algebra is itself a consequence of the underlying physics. For completeness, Modular Composite Representation (MCR)~\cite{snaider2014mcr} is best understood as a representational model rather than a hardware category; concrete MCR implementations can certainly be efficient, but the representational choice itself does not establish an intrinsic physical algebra.

\paragraph{Physics-native subroutines and partial primitive alignment.}
Optical correlators and holographic processors based on Fourier optics~\cite{goodman2005} implement correlation and convolution through diffraction and lens-based transforms. They demonstrate that convolution-like operations can be realized in effectively constant physical time, but they typically rely on spatial holography and tight mechanical tolerances, which limits robustness and integration.

Spin-wave (magnonic) devices leverage spin waves in magnetic materials for interference-based logic and neuromorphic computing~\cite{menezes2024magnonic}. These systems show that wave-based substrates beyond electromagnetics can implement logic-like operations, but they do not yet amount to an end-to-end HDC/VSA realization.

RF correlators and matched filters form the backbone of spread-spectrum communications and radar~\cite{proakis2007,pozar2012}. They physically compute inner products between waveforms and thus realize similarity operations directly at the signal-processing level. However, they do not provide a complete set of HDC primitives nor a mathematically invertible embedding of a high-dimensional symbolic space.

The in-memory factorizer of Langenegger et al.~\cite{langenegger2023factorizer} moves closer to native physics by exploiting both the computation-in-superposition of holographic representations and the intrinsic stochasticity of analogue phase-change-memory hardware, and experimentally demonstrates large-scale factorization on two in-memory compute chips. A recent quantum-native proposal, QHDC~\cite{cumbo2025qhdc}, maps bundling, binding, permutation, and similarity to native quantum procedures and reports execution on an IBM Heron processor. These works are important waypoints because they show that specific HDC-relevant operations can align naturally with substrate physics. However, they are best read as demonstrations of native subroutines, task-specific factorization, or operator mappings, rather than as end-to-end evidence that the HDC/VSA algebra is intrinsically revealed by the substrate without an externally imposed representational layer.

\paragraph{Physics-based instantiation: Resonant Field Computing.}
Against this background, the claim advanced in the present paper is narrower and sharper than a generic accelerator claim. The Unitary Wave Embedding maps discrete bipolar hypervectors to broadband coherent waveforms using a unitary transform. Under this map,
\begin{itemize}
    \item bundling becomes linear superposition of fields;
    \item permutation becomes coherent phase evolution;
    \item binding is realized as nonlinear spectral mixing combined with engineered aliasing that enforces a circular-convolution structure; and
    \item similarity is proportional to a measurable differential power $\Delta P$ derived from the quadratic field functional.
\end{itemize}
These properties arise from the chosen unitary embedding and the native dynamics of coherent wavefields, rather than from per-component symbolic operations executed by an external digital controller. Because binding presently requires nonlinear mixing together with an engineered aliasing/projection step, the strongest accurate framing is that RFC is a \emph{candidate} algebra-revealing physical instantiation of HDC/VSA. What is established in this manuscript is a connected RFC/UWE stack: the embedding preserves geometry, the primitive correspondences are made explicit, the binding pipeline reproduces the discrete operation under the stated aliasing plan, and the similarity/readout layer is tied to measurable differential power and validated at array scale in full-wave simulation; the architecture sections then delimit the conditions under which a scalable hardware realization would need to operate.

\section{Methods}
This section is organized to be read in three passes. Section~3.1 defines the Unitary Wave Embedding (UWE), Section~3.2 maps each HDC/VSA primitive to its physical counterpart, and Section~3.3 shows how similarity is recovered from measurable power. Throughout, UWE denotes the mathematical embedding, whereas RFC denotes the full physical architecture and readout protocol built on that embedding.

\paragraph{Notation used in this section.}
I write a discrete hypervector as \(x\in\mathbb{R}^N\) (bipolar \(x_i\in\{\pm1\}\) in the numerical experiments), its unitary-DFT coefficients as \(X=Fx\), and its embedded waveform as \(s_x(t)=U(x)(t)\). I use \(\odot\) for component-wise multiplication in vector space, \(\star\) for circular convolution over frequency-bin index, and \(*\) for linear convolution in continuous frequency. Inner products \(\langle\cdot,\cdot\rangle\) are taken in the space implied by context.

\subsection{Unitary Wave Embedding (UWE)}
UWE is a change of coordinates from a finite-dimensional vector space to a finite-dimensional subspace of waveforms. The key requirement is that this map is unitary (energy- and angle-preserving), so similarity in vector space is preserved in waveform space.

Although I present the construction using bipolar vectors because that is
the baseline HDC/VSA setting in this paper, the same map extends to
real-valued \(x\in\mathbb{R}^N\); for complex vectors, the derivation is
unchanged once the real-valued output constraint is relaxed. The numerical
validation in this draft is presented on the bipolar MAP-B baseline.

Let \(F\in\mathbb{C}^{N\times N}\) be the unitary DFT with \(1/\sqrt{N}\) normalization, so \(F^*F=I\) \citep{oppenheim1999}. Choose an orthonormal tone set \(\{\phi_k\}_{k=0}^{N-1}\) on a finite window \([0,T]\):
\[
\phi_k(t) \;=\; \frac{1}{\sqrt{T}}\,e^{j2\pi f_k t},
\qquad
\int_0^T \phi_k(t)\,\phi_\ell^*(t)\,dt \;=\; \delta_{k\ell}.
\]
To ensure orthogonality over \([0,T]\), use a symmetric comb
\[
f_k \;=\; f_{\mathrm{cen}} + \Big(k-\tfrac{N-1}{2}\Big)\,\Delta f,
\qquad \Delta f \;=\; \frac{1}{T}.
\]
Define the embedding \(U:\mathbb{C}^N\to L^2[0,T]\) by
\[
U(x)(t) \;=\; \sum_{k=0}^{N-1} (Fx)_k\,\phi_k(t).
\]
For real-valued waveform synthesis, impose Hermitian symmetry on DFT coefficients, \((Fx)_{N-k}=\overline{(Fx)_k}\), so that \(U(x)\) is real-valued. By Parseval/Plancherel, inner products and norms are preserved \citep{oppenheim1999,stein2003}:
\[
\langle Ux,Uy\rangle_{L^2} \;=\; \langle x,y\rangle,
\qquad
\|Ux\|_{L^2}=\|x\|_2.
\]

\begin{theorem}[Power--Similarity Bridge]
For any \(x,y\in\mathbb{R}^N\) and the unitary embedding \(U\) above,
\[
\Delta E(x,y) := \|U(x)+U(y)\|_{L^2}^2 - \|U(x)\|_{L^2}^2 - \|U(y)\|_{L^2}^2
= 2\langle x,y\rangle.
\]
If a physical readout maps this interference quantity to differential power with calibration factor \(\kappa\), then
\[
\Delta P_{\mathrm{phys}} = \kappa\,\Delta E = 2\kappa\langle x,y\rangle.
\]
For normalized vectors, \(\Delta P_{\mathrm{phys}} = 2\kappa\cos\theta\).
\end{theorem}
\paragraph{Proof sketch.}
Apply the polarization identity \(\|a+b\|^2-\|a\|^2-\|b\|^2=2\langle a,b\rangle\) with \(a=U(x)\), \(b=U(y)\), then use the UWE isometry \(\langle U(x),U(y)\rangle=\langle x,y\rangle\) (Plancherel; \citep{stein2003,oppenheim1999}). The calibration factor \(\kappa\) then collects geometry, impedance, coupling, and instrumentation factors that map the ideal interference quantity to measured physical power units.
\noindent\textit{Interpretation.} The exact Hilbert-space identity is an interference-energy statement; once \(\kappa\) is calibrated, the corresponding differential power measurement directly reports cosine similarity.

\subsection{RFC Primitives}
With UWE in place, each HDC/VSA primitive can be read as a ``discrete operation $\rightarrow$ physical operation'' mapping.

\paragraph{Bundling (superposition).}
In HDC/VSA, bundling is vector addition (often followed by thresholding/sign if a bipolar representation is required). Under UWE, this is simply field superposition:
\begin{equation}
s_{\text{bundle}}(t) = s_A(t) + s_B(t).
\end{equation}
For quasi-orthogonal source vectors, Parseval’s identity implies that total energy is approximately additive \citep{oppenheim1999}, so the bundle remains similar to its constituents.

\paragraph{Binding (nonlinear realization with spectral wrapping).}
In vector space, bipolar binding is component-wise multiplication. Under UWE, naive waveform multiplication produces linear spectral convolution rather than the required circular convolution, so an engineered wrapping/aliasing step is introduced to restore the discrete target operation.
\textbf{Discrete target.} In bipolar HDC/VSA, binding is component-wise multiplication \(z = x \odot y\). In the spectral domain this is circular convolution:
\begin{equation}
  Z_k
  = (X \star Y)_k
  := \sum_{m=0}^{N-1} X_m \, Y_{(k - m)\bmod N},
  \label{eq:spectral-binding-exact}
\end{equation}
which is the standard modulo-\(N\) convolution rule over frequency-bin index.

\textbf{Stage 1: nonlinear mixing.} The first physical step multiplies waveforms in time (classical product-to-sum / prosthaphaeresis):
\begin{equation}
  s_z(t) = s_x(t)\,s_y(t).
\end{equation}
This is implemented by a nonlinear device (e.g., a mixer) and generates sum and difference frequency components \citep{pozar2012}. In frequency, this yields a \emph{linear} convolution,
\begin{equation}
  Z_{\mathrm{lin}}(\omega) = (S_x * S_y)(\omega),
\end{equation}
whose support is roughly twice the original comb bandwidth, so it does not yet match the discrete circular rule.

\textbf{Stage 2: spectral wrapping (engineered aliasing).} To recover circular structure, the expanded linear spectrum is folded back into the original \(N\)-bin band. Let \(B\) denote the UWE comb bandwidth. Periodization gives
\begin{equation}
  Z_{\mathrm{circ}}(\omega)
  \;=\;
  \sum_{m=-\infty}^{\infty}
  Z_{\mathrm{lin}}(\omega - m B).
  \label{eq:Z-circ-periodization}
\end{equation}
Equivalently, convolve with a Dirac comb:
\begin{equation}
  \chi_{\mathrm{wrap}}(\omega)
  \;=\;
  \sum_{m=-\infty}^{\infty} \delta(\omega - m B),
  \qquad
  Z_{\mathrm{circ}}(\omega) = \bigl(Z_{\mathrm{lin}} * \chi_{\mathrm{wrap}}\bigr)(\omega).
  \label{eq:chi-wrap-continuous}
\end{equation}
By the convolution theorem, this corresponds in time to multiplication by a sampling train,
\begin{equation}
  \mathcal{F}^{-1}\{\chi_{\mathrm{wrap}}\}(t)
  =
  \sum_{n=-\infty}^{\infty} \delta(t - n T_s),
  \qquad T_s = \frac{1}{B},
\end{equation}
so
\begin{equation}
  z_{\mathrm{wrapped}}(t)
  =
  z_{\mathrm{lin}}(t)\,
  \sum_{n=-\infty}^{\infty} \delta(t - n T_s).
\end{equation}
Intuitively, sampling at the comb-matched rate folds the out-of-band ``linear tail'' back into baseband, recreating modulo-\(N\) accumulation.

In a discretized implementation with \(N\) bins spaced by \(\Delta f\), the wrapped spectrum is
\begin{equation}
  Z_k^{\mathrm{phys}}
  \;\approx\;
  \sum_{m=0}^{N-1}\sum_{n=0}^{N-1}
    X_m \, Y_n \, \chi_{\mathrm{wrap}}(m,n,k),
  \label{eq:spectral-binding-phys}
\end{equation}
with ideal aliasing kernel
\begin{equation}
  \chi_{\mathrm{wrap}}(m,n,k)
  :=
  \begin{cases}
    1, & n \equiv (k - m) \pmod N,\\[4pt]
    0, & \text{otherwise}.
  \end{cases}
  \label{eq:chi-wrap-ideal}
\end{equation}
which reduces \eqref{eq:spectral-binding-phys} exactly to \eqref{eq:spectral-binding-exact}.

In the numerical RFC binding experiments (Appendix~\ref{app:binding}), this is implemented as explicit modulo-\(N\) frequency re-binning: after low-pass isolation of the difference band, each FFT bin at frequency \(f_i\) is mapped to \(k=(\mathrm{round}(f_i/\Delta f))\bmod N\) and accumulated into \(Z_k\). This is the discrete realization of \(\chi_{\mathrm{wrap}}\). Appendix~\ref{app:binding} verifies that this pipeline reproduces discrete binding up to numerical precision.

\paragraph{Permutation (cyclic shift as phase evolution/time shift).}
In HDC/VSA, permutation reorders indices, usually by cyclic shift. By the Fourier shift theorem \citep{oppenheim1999}, that discrete shift is a linear phase ramp in frequency, which appears physically as a time delay:
\begin{equation}
s(t) \mapsto s(t - \tau).
\end{equation}
This operation preserves norm and the quasi-orthogonality structure used by role/filler encoding. Appendix~\ref{app:permutation} provides a direct numerical equivalence test between discrete cyclic shifting and waveform time delay.

\subsection{Similarity, Interference Energy, and Differential Power}
Similarity readout is based on a quadratic measurement and can be understood as a three-measurement protocol: measure each self-power, then measure the combined waveform.

For two signals \(s_x(t)\) and \(s_y(t)\), total superposition power is
\begin{equation}
P_\Sigma = P_x + P_y + 2 \langle s_x, s_y \rangle,
\end{equation}
where \(\langle s_x, s_y \rangle\) is time-averaged cross-correlation (the standard signal-similarity quantity in DSP~\citep{proakis2007}). Subtracting self-powers isolates the ideal interference term:
\begin{equation}
\Delta E = P_\Sigma - (P_x + P_y) = 2 \langle s_x, s_y \rangle.
\end{equation}
Under UWE, Plancherel gives \(\Delta E = 2\langle x,y\rangle\). In the measured physical system, this interference quantity is reported as a differential power. In the idealized uncoupled model,
\begin{equation}
\Delta P_{\mathrm{phys}} \approx \kappa\,\Delta E = 2\kappa \cdot \cos\theta,
\end{equation}
with fixed calibration constant \(\kappa\) that absorbs impedance, geometry, coupling, and instrumentation scale factors.

For retrieval, each stored vector is assigned to a physical \emph{library emitter} \(L_j\). The hardware readout at channel \(j\) is the differential power between a query-on run and a baseline run:
\begin{equation}
  \Delta P_j = P^{(\mathrm{query})}_j - P^{(\mathrm{baseline})}_j,
\end{equation}
optionally normalized by a reference isolated-emitter power \(P_{\mathrm{ref}}\):
\begin{equation}
  \Delta \tilde P_j = \frac{\Delta P_j}{P_{\mathrm{ref}}}.
\end{equation}
The score \(\Delta \tilde P_j\) is the per-channel retrieval quantity read by a lock-in detector and, by the Power--Similarity Bridge, is linear in cosine similarity between the query and the \(j\)-th library hypervector.

At the array-readout level, this model is validated by 3D FDTD simulations with dipole emitters. Match discriminability is summarized by the Correlation Contrast Ratio (CCR):
\begin{equation}
\mathrm{CCR} =
\frac{\big|\Delta P_{\mathrm{match}} - \Delta P_{\mathrm{non-match}}\big|}
     {P_{\mathrm{baseline,match}} + P_{\mathrm{baseline,non-match}}}.
\end{equation}
CCR is a global test metric (not the per-channel retrieval score \(\Delta \tilde P_j\)). In the documented \(N=1000\) coupled calibration, the simulation gives \(\mathrm{CCR}_{\mathrm{cpl}}\approx 8.7\times10^{-5}\). Although small, this is consistent in sign and order with the analytical model (\(\mathrm{CCR}\approx10^{-3}\) under idealized Hertzian-dipole assumptions), supporting a physical (non-artifactual) interaction effect.

\paragraph{Coupled (saturated) versus emitter-isolated CCR.}
The CCR reported above is for a fully mutually coupled array (no emitter isolation or mantle cloaks), i.e., a saturated comparison architecture. I denote it \(\mathrm{CCR}_{\mathrm{cpl}}\), where the denominator baseline includes substantial library--library coupling contributions.

In the emitter-isolated architecture (Section~\ref{sec:emitter_isolation}), the baseline at each library emitter is instead its isolated self-power, \(P_{\mathrm{iso}}(L_j)\), motivating
\[
\mathrm{CCR}_{\mathrm{iso}} =
\frac{\big|\Delta P(L_{\mathrm{match}},Q) - \Delta P(L_{\mathrm{non-match}},Q)\big|}
     {P_{\mathrm{iso}}(L_{\mathrm{match}}) + P_{\mathrm{iso}}(L_{\mathrm{non-match}})}.
\]
This uses the same Power--Similarity mechanism, but with a different physical baseline once emitter--emitter coupling is suppressed.

\section{Discussion}
\subsection{On the Nature of the Wave–Geometric Duality}
The theoretical framework presented here argues that key parts of the abstract algebra of HDC can be read directly from the physics of coherent linear and weakly nonlinear fields once the UWE is adopted. The bundling primitive (\textit{addition}) is physically realized as wave superposition, permutation (\textit{cyclic shift}) maps directly to phase evolution, and similarity is recovered from a quadratic power functional. Binding (\textit{multiplication}) is reproduced through nonlinear spectral mixing together with the explicit aliasing/projection step developed in this manuscript. Accordingly, the most precise statement is not that raw electromagnetics automatically implements every HDC primitive, but that an engineered wave substrate can be made algebraically equivalent to HDC/VSA under a specified embedding and readout protocol. In that sense, RFC is framed here as a candidate algebra-revealing physical instantiation rather than merely a faster emulator.

This duality should be read as \textbf{natural in principle but engineered in realization}. The Fourier/convolution identities are fundamental; turning them into a robust computing system is not. Practical RFC requires controlled coherence, calibrated spectral structure, and isolation of unwanted emitter--emitter coupling so that the \(\Delta P\) cross term remains measurable. In that sense, the conceptual result is a physics-level correspondence, while the bottleneck is instrumentation and array engineering. This perspective is aligned with substrate-first computing ideas: computation is extracted from the native dynamics of the medium, then made reliable through explicit design constraints.

\subsection{Binding Constraints and Frequency Centering}
The duality between discrete binding (component-wise (Hadamard) multiplication) and physical mixing is exact only under specific spectral conditions. Component-wise (Hadamard) multiplication in the vector domain corresponds to \textit{circular} convolution in the spectral domain. However, the physical mixing of two waveforms, $s_z(t) = s_x(t) \cdot s_y(t)$, produces a \textit{linear} convolution of their spectra. The mathematical equivalence to HDC binding is broken unless this linear convolution can be made to replicate a circular one.

This leads to two primary constraints. First, \textbf{modulo-N aliasing}: for the linear convolution to approximate the circular convolution, the resulting sum and difference frequencies must "wrap around" the spectral band. Second, \textbf{spectral placement and bookkeeping}: the abstract derivation is cleanest on a cyclic comb written in centered form, while practical real-signal simulations often use a positive-frequency or translated-passband parametrization. The key requirement is not a unique absolute carrier placement, but that the tone spacing, coherent phase reference, filter band, and wrapping rule all be defined consistently within the chosen convention. Real implementations, which operate at RF frequencies (e.g., in the MHz or GHz range), therefore require an engineered projection, aliasing, to restore the modulo-$N$ spectral structure required for equivalence to discrete binding under the UWE.

At the level of the UWE model, this aliasing requirement is satisfied in the numerical binding experiment of Appendix~\ref{app:binding}, where the wrapped spectrum recovers the target bound vector with unit cosine similarity.

\begin{figure}[h]
    \centering
    \includegraphics[width=0.5\linewidth]{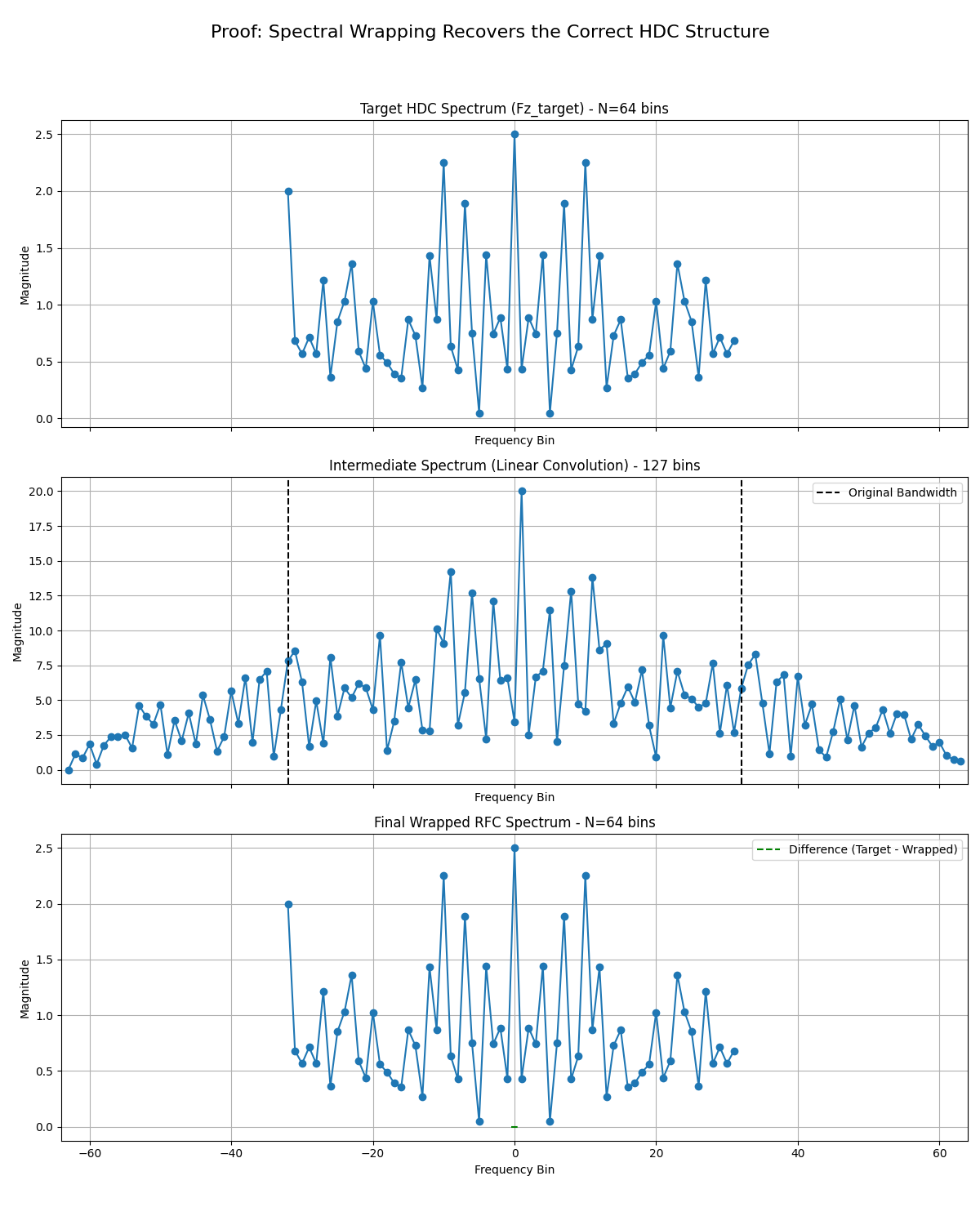}
    \caption{Binding pathways compared. Left: target HDC circular-convolution structure. Center: raw UWE prosthaphaeresis output (linear-convolution expansion). Right: engineered aliasing/wrapping that restores modulo-\(N\) structure.}
    \label{fig:binding_aliasing}
\end{figure}

Interestingly, the challenge of physically implementing the binding operation is a recurring theme in vector symbolic architectures. It is plausible that phase-coherent binding, as seen in Holographic Reduced Representations (HRR) and related models, may arise naturally in embeddings that leverage phase information more directly. Figure~\ref{fig:binding_aliasing} contrasts the target circular-convolution structure with the raw mixed spectrum and the wrapped spectrum used in RFC.

\subsection{Theoretical Throughput Bounds}
A primary advantage of RFC is \textbf{parallel scaling with library size}: under the parallel-wave interaction model, the query field interacts with all library emitters simultaneously, so the physical interaction step does not grow linearly with library size \(L\).

To compare this physical process with a digital baseline, an \textbf{equivalent-FLOPs} proxy can be used. One RFC query corresponds to \(L\) parallel dot products, each requiring \((2N-1)\) floating-point operations, giving
\begin{equation}
    \text{Equivalent GFLOPS} = \frac{L \times (2N - 1)}{\text{Latency (s)} \times 10^9}.
\end{equation}
This proxy highlights computational density for large-scale similarity search.

End-to-end latency is nevertheless not strictly independent of vector dimensionality \(N\), because the observation window is tied to comb spacing (\(\Delta f = 1/T\)) and to the readout sensitivity required to resolve small \(\Delta P\) perturbations. When mapped to GHz carriers, the physical interaction window can still be nanosecond-scale, but practical throughput remains measurement-limited rather than propagation-limited. In the documented coupled-array calibration, \(\mathrm{CCR}_{\mathrm{cpl}}\approx 8.7\times10^{-5}\), implying the match/non-match swing is far below the baseline scale. Isolation mechanisms (mantle-cloak targets or validated surrogate isolators) improve this operating point by reducing coupling-dominated baseline power; cloak-specific latency gains are left as future engineering validation.

\subsection{Robustness and Graceful Degradation}
A primary motivation for hyperdimensional computing is robustness to noise and component failure. For the wave--geometric duality to be a viable computational framework, it must replicate both the HDC algebra and this resilience. This is evaluated with two numerical stress tests targeting distinct error sources: physical noise in the continuous wave domain and representational noise in the discrete vector domain.

The first test injects Additive White Gaussian Noise (AWGN) into waveforms \(s_x(t)\) and \(s_y(t)\) prior to multiplication, modeling thermal noise or channel interference \citep{proakis2007}. The second test introduces graceful-degradation stress by randomly flipping a percentage of source-vector components in \(x\) and \(y\) before UWE synthesis, modeling storage/representation errors \citep{kanerva2009,kleyko2022survey}.

The results, summarized in Table~\ref{tab:noise_robustness}, demonstrate exceptional resilience.

\begin{table}[h!]
 \caption{Noise Robustness of the RFC Binding and Unbinding Process.}
 \label{tab:noise_robustness}
 \centering
 \begin{tabular}{l c c c}
    \toprule
    \multicolumn{2}{l}{\textbf{A: Physical Noise (AWGN on Waveforms)}} & \multicolumn{2}{c}{\textbf{Unbound Vector Recovery}} \\
    \cmidrule(r){1-2} \cmidrule(l){3-4}
    \textbf{SNR} & \textbf{Bound} \(\cos(z, \hat{z})\) & \textbf{Unbound} \(\cos(y, \hat{y})\) & \textbf{Sign Accuracy (\%)} \\
    \midrule
    20 dB & 0.9999 & 0.9999 & 100.00 \\
    10 dB & 0.9994 & 0.9994 & 100.00 \\
    0 dB  & 0.9931 & 0.9931 & 100.00 \\
    \midrule
    \multicolumn{2}{l}{\textbf{B: Representational Noise (Bit-flips)}} & \multicolumn{2}{c}{\textbf{Unbound Vector Recovery}} \\
    \cmidrule(r){1-2} \cmidrule(l){3-4}
    \textbf{Flip Prob.} & \textbf{Bound} \(\cos(z, \hat{z})\) & \textbf{Unbound} \(\cos(y, \hat{y})\) & \textbf{Sign Accuracy (\%)} \\
    \midrule
    1.0\%  & 0.9531 & 0.9531 & 97.66 \\
    10.0\% & 0.7062 & 0.7062 & 85.31 \\
    20.0\% & 0.2813 & 0.2813 & 64.06 \\
    \bottomrule
  \end{tabular}
\end{table}

The architecture exhibits remarkable resilience to physical noise. Even at a signal-to-noise ratio of 0~dB, where noise power equals signal power, the recovered vectors retain perfect sign accuracy and a cosine similarity greater than 0.99. This demonstrates that information is encoded in the collective, coherent structure of the waveform. The distributed nature of the UWE provides a significant processing gain against incoherent noise, analogous to spread-spectrum systems \citep{proakis2007}. 

Furthermore, the system displays the canonical graceful degradation of HDC \citep{kanerva2009,kleyko2022survey}. When 10\% of the source vector components are corrupted, the recovered vector is still highly correlated with the original (cosine similarity \(\approx 0.7\)), making it far closer to the correct item than to any random vector in the space. This confirms that the UWE and its associated physical operations faithfully preserve the metric space of the parent hyperspace. These numerical results provide strong evidence that the RFC framework is not a fragile theoretical construct but a robust computational model that truly inherits the most vital properties of HDC.

\subsection{Outlook}
This framework lays the groundwork for a new class of analog co-processors for brain-inspired AI. The immediate next step is the fabrication and testing of a small-scale RFC prototype to validate the simulation results and refine the engineering design. Future research should also explore scaling the architecture to larger library sizes and investigating the use of different wave modalities, such as acoustics or guided optical waves. Furthermore, the wave-centric perspective may enable the development of entirely new computational primitives beyond the standard HDC set, potentially leading to more powerful and efficient algorithms for complex cognitive tasks. Ultimately, RFC could serve as a specialized, high-throughput co-processor for large-scale similarity search, a foundational operation in many modern AI systems.

Because the Unitary Wave Embedding is linear, invertible, and norm-preserving, and because similarity can be recovered from a single quadratic measurement, the framework may extend beyond hyperdimensional computing. In particular, any data that can be embedded into high-dimensional vectors could, in principle, be mapped into wavefields and processed via RFC-style interactions. Exploring connections to continuous attention mechanisms and more general tensor operations is an open direction for future work; the present paper establishes only the building blocks required for such investigations.

\section{Physical Constraints and Validity Conditions}
\subsection{Compositionality}
Routing of signals enabling compositionality is expected to work in one of two ways: (i) ADC/DAC capture and replay of hypervectors with distortion kept within the robustness margins demonstrated in §4.4, or (ii) future realizations of full wave-domain logic that preserve the coherence and tone structure required by the UWE. The design of such routing and control mechanisms is an open engineering problem and is not addressed in this work.

\subsection{Phase-Locking Assumption.}
Throughout this work all waveforms participating in a given RFC operation
are assumed to be mutually phase-locked; that is, their constituent tones
share a common frequency comb $\{f_k\}$ and phase reference.
This ensures that the embedding $U$ remains unitary
and that inner products, superpositions, and time delays
are evaluated within a single coherent frame.

\subsection{Phase Jitter Tolerance of RFC Binding}
\label{sec:jitter-binding}

The wave--geometric duality and the power--similarity bridge rely on coherence: the embedded spectra must share a common phase reference so that the isometry
\(\langle Ux,Uy\rangle_{L^2}=\langle x,y\rangle\) is approximately preserved.
In hardware, the dominant threat is not a global (common-mode) phase rotation of the entire comb, but \emph{relative} phase errors across tones or across signal paths, which distort the embedded waveform shape.

I model a conservative impairment as \textbf{independent per-tone phase jitter} applied at the embedding stage:
\begin{equation}
\widetilde{X}_k = X_k e^{j\delta\phi_k}, \qquad
\delta\phi_k \sim \mathcal{N}(0,\sigma_\phi^2),
\end{equation}
(and similarly for \(\widetilde{Y}_k\)). This is pessimistic compared to many physical oscillators where phase noise is substantially correlated across tones; nevertheless it provides a useful worst-case tolerance envelope. Additional details, including the injection model and metrics, are given in Appendix~\ref{app:jitter}.

\paragraph{Measured degradation (2D FDTD binding test).}
Using the FDTD-based binding pipeline (prosthaphaeresis + low-pass isolation of the difference band + modulo-\(N\) spectral wrapping), I swept \(\sigma_\phi\) and measured (i) cosine similarity between the recovered bound vector and the ideal discrete binding \(z=x\odot y\), and (ii) sign (bit) accuracy after binarization. Table~\ref{tab:jitter_binding} summarizes representative results.

\begin{table}[h!]
\centering
\caption{Phase-jitter tolerance of RFC binding under independent per-tone Gaussian phase perturbations (representative run).}
\label{tab:jitter_binding}
\begin{tabular}{c c c}
\toprule
\(\sigma_\phi\) (rad) & \(\cos(z_{\mathrm{target}}, \hat{z})\) & Sign Accuracy (\%) \\
\midrule
0.00 & 0.9987 & 100.00 \\
0.10 & 0.9942 & 100.00 \\
0.20 & 0.9782 & 100.00 \\
0.50 & 0.8539 & 96.88 \\
1.00 & 0.4378 & 67.19 \\
\bottomrule
\end{tabular}
\end{table}

Two regimes are apparent. For \(\sigma_\phi \le 0.2\) rad (\(\approx 11.5^\circ\) RMS), the binding operator remains essentially ideal at the HDC level (cosine similarity \(\ge 0.97\) with perfect sign recovery). Between \(\sigma_\phi \approx 0.2\) and \(0.5\) rad, performance degrades but remains strongly correlated with the ideal bound vector, consistent with canonical HDC graceful degradation \citep{kanerva2009,kleyko2022survey}. At \(\sigma_\phi=1.0\) rad (\(\approx 57^\circ\) RMS), binding becomes substantially noisy but remains above chance (sign accuracy \(>50\%\)), meaning a clean-up memory could still recover correct items in many VSA workflows.

\paragraph{Implications for the \(\Delta P\) similarity mechanism.}
Relative phase jitter also attenuates the coherent cross term that produces \(\Delta P\). Under a simple independent-phase-error model, the expected coherent contribution is reduced by the phasor mean
\(\mathbb{E}[e^{j\delta\phi}]=e^{-\sigma_\phi^2/2}\), implying an approximate scaling
\begin{equation}
\mathbb{E}[\Delta P] \;\approx\; 2\kappa\,e^{-\sigma_\phi^2/2}\,\langle x,y\rangle,
\end{equation}
so that jitter reduces the match/non-match swing and can collapse CCR if \(\sigma_\phi\) becomes large. This motivates treating \(\sigma_\phi\) (or an equivalent clock-jitter budget) as a first-class design constraint for RFC co-processors.

\paragraph{Relation to timing jitter.}
Timing error \(\delta\tau\) in permutation or sampling introduces a frequency-dependent phase error \(\delta\phi_k \approx 2\pi f_k \delta\tau\), so wideband combs translate even small \(\delta\tau\) into significant per-tone phase error at high \(f_k\) \citep{oppenheim1999,proakis2007}. The envelope in Table~\ref{tab:jitter_binding} therefore also provides an interpretable \emph{engineering} target: maintain relative per-tone RMS phase error \(\sigma_\phi \le 0.2\) rad for near-ideal binding, with useful operation persisting into the \(\sim 0.5\) rad regime due to HDC robustness.

\subsection{Emitter Isolation as an Implementation Requirement}
\label{sec:emitter_isolation}

\paragraph{Why isolation is required.}
The $\Delta P$ similarity mechanism is present in the fully mutually coupled array, but the measured baseline at each
library emitter includes substantial library--library near- and mutual-field coupling. This creates a saturated
comparison regime in which the match/non-match swing in $\Delta P$ is a small perturbation on a coupling-dominated baseline,
limiting practical readout speed and stability.

\paragraph{Mantle cloaking requirement (primary realization).}
The intended resolution is an emitter-isolated library in which each library radiator is encased in an anisotropic conformal
mantle cloak designed to suppress its scattered field into in-plane modes that dominate inter-library coupling,
while remaining responsive to an off-plane incident query field. In this emitter-isolated regime, the baseline scale at each
library emitter approaches its isolated self-power $P_{\mathrm{iso}}(L_j)$, and the appropriate figure of merit is the isolated CCR
defined in Section~3.3. This is an engineering requirement imposed by the architecture; this paper does not claim mantle-cloak
performance from metasurface simulations.

\paragraph{Surrogate isolation validation (controlled anisotropic isolator).}
To validate the architectural premise that anisotropic emitter isolation lifts saturation, I performed a controlled 2D surrogate
experiment in which an anisotropic isolator is implemented by a lossy baffle placed only on the in-plane library--library path
(thereby suppressing L--L coupling while leaving the off-plane query path open). In this surrogate geometry the coupling-dominated
baseline scale decreased by $32.67$~dB and the resulting contrast ratio increased by approximately $241\times$ relative to the
unisolated surrogate geometry, while match and mismatch queries continued to yield distinct $\Delta P$ values.
Full details of this surrogate experiment (geometry, measurement protocol, and numerical values) are provided in
Appendix~\ref{app:isolation_surrogate}.

\subsection{Requirements for Physical Realization}
This paper is intentionally technology-agnostic: the results do not depend on whether the carrier is RF, mmWave, or photonic, provided the implementation satisfies the coherence, spectral, and measurement conditions implied by the Unitary Wave Embedding (UWE) and the Power--Similarity bridge. What follows is therefore a requirements-level specification, not a device proposal.

\paragraph{(1) Mutually coherent tone-comb generation and distribution.}
RFC relies on the coherent cross term that produces $\Delta P$; this term collapses if the constituent tones do not share a common phase reference. A practical implementation therefore requires (i) a multi-tone source (comb) whose tones are mutually phase-locked, and (ii) a distribution network that preserves relative phase across all library channels and the query channel to within the tolerance quantified in Section~\ref{sec:jitter-binding}. In RF this suggests a shared LO/clock tree and calibrated path delays; in photonics it suggests a shared comb source and stable interferometric routing. Common-mode phase rotation is not the dominant issue; differential phase error across tones and across channels is.

\paragraph{(2) Library/query field geometry consistent with the comparison model.}
Within the broader RFC/UWE stack, the similarity/readout mechanism is derived under a superposition model in which the query field interacts with each library emitter and the library emitter's measured power changes by a small correlation-dependent amount. The 3D FDTD results in this paper validate that this mechanism persists even in a fully mutually coupled array, but they also show that the \emph{baseline} power at each library emitter can be dominated by library--library coupling in dense arrays. Consequently, a hardware realization must explicitly manage geometry and isolation to prevent saturation of the readout (Section~\ref{sec:emitter_isolation}).

\paragraph{(3) Emitter isolation to lift saturation (architecture-level requirement).}
This work distinguishes a coupled (saturated) regime from an emitter-isolated regime via $\mathrm{CCR}_{\mathrm{cpl}}$ versus $\mathrm{CCR}_{\mathrm{iso}}$. The intended physical realization of emitter isolation is an anisotropic mantle cloak per library emitter that suppresses in-plane scattering/coupling pathways while remaining responsive to an off-plane query field (Section~\ref{sec:emitter_isolation}). This paper does not claim mantle-cloak performance from cloak simulations; rather, it states an isolation target (e.g., $\ge 40$~dB across the operating comb) as an engineering requirement implied by the readout physics. To validate the architecture premise independently of cloak-specific electromagnetics, Appendix~\ref{app:isolation_surrogate} reports a controlled surrogate isolator (a lossy anisotropic baffle) that suppresses the in-plane library--library path while leaving the query path open, yielding a large baseline reduction and a corresponding increase in contrast. In hardware terms, mantle cloaking is the preferred low-loss, conformal route; surrogate isolators (e.g., recessing elements into absorptive cavities or baffles) are credible prototyping mechanisms to demonstrate the saturation-to-isolated transition even before metasurface optimization.

\paragraph{(4) Measurement chain and dynamic range.}
The observable is a differential power $\Delta P_j = P^{(\mathrm{query})}_j - P^{(\mathrm{baseline})}_j$ measured at each library emitter (Section~3.3). In the coupled regime the CCR can be small (validated here by 3D FDTD), so the measurement chain must resolve a correlation-dependent perturbation on top of a potentially large baseline. Practically this points to (i) phase-sensitive detection / lock-in style readout keyed to the query modulation or an explicit baseline/query differencing protocol, (ii) careful control of leakage paths that would masquerade as $\Delta P$, and (iii) calibration against a reference power scale (the role of $P_{\mathrm{ref}}$ in the simulations). The specific electronics are implementation-dependent, but the requirement is not: the readout must reliably resolve a small coherent cross term in the presence of a larger incoherent (or coupling-dominated) background.

\paragraph{(5) Nonlinear device for physical binding (when binding is performed in-wave).}
If binding is executed physically (rather than digitally), the architecture requires a controlled nonlinear mixing device to implement prosthaphaeresis, followed by an engineered aliasing/wrapping stage that enforces the modulo-$N$ spectral structure (Appendix~\ref{app:binding}). This can be implemented as explicit sampling/decimation matched to the comb spacing, or an equivalent spectral re-binning mechanism. If binding is executed digitally and only similarity is executed physically, the nonlinear requirement can be relaxed; the coherence and measurement requirements remain.

\paragraph{(6) Routing for compositionality.}
Section~\ref{sec:emitter_isolation} and the compositionality constraint emphasize that full system-level HDC workflows require routing of intermediate representations. In the near term this plausibly involves ADC/DAC capture and replay, with distortion bounded by the robustness margins demonstrated in the noise/jitter experiments. A fully wave-domain routing fabric that preserves UWE coherence is an open engineering problem and is not claimed here.

In summary, the hardware message of this paper is precise but narrow: the UWE together with the primitive and readout mappings specify what a physical RFC co-processor must satisfy (coherence, frequency planning, isolation, and measurement sensitivity). The 3D coupled-array simulations demonstrate that the correlation effect is physically present even under strong mutual coupling, while the binding study, isolation section, and surrogate appendix specify the additional conditions needed for a scalable stack.

\section{Conclusion}
This work establishes a wave--geometric duality for hyperdimensional computing by constructing an explicitly unitary embedding (UWE) that maps bipolar hypervectors into coherent broadband waveforms while preserving inner products. Under this embedding, the HDC/VSA primitives align with wave physics: bundling becomes linear superposition, permutation corresponds to coherent phase evolution (time shift), and binding admits a physical realization via nonlinear spectral mixing combined with an engineered spectral wrapping/aliasing kernel that reproduces circular convolution structure.

The similarity/readout layer then supplies a measurable bridge between geometry and hardware: the correlation term of a quadratic power measurement yields a calibrated differential power that is linearly proportional to the hypervector inner product (and therefore to cosine similarity for normalized vectors). This converts similarity retrieval from an algorithmic dot product into a physically instantiated interference measurement, provided the system maintains coherence and frequency structure. In this sense similarity receives the deepest treatment in the paper not as a replacement for the rest of the stack, but because it determines how the broader RFC/UWE program becomes measurable and what constraints the architecture must satisfy. The paper also states explicit validity conditions, including phase-locking and jitter tolerance bounds, and demonstrates that the binding pipeline reproduces discrete HDC behavior under the specified aliasing plan.

Full-wave FDTD studies corroborate the physically nontrivial parts of this framework. In the documented $N=1000$ coupled calibration, the array is fully mutually coupled and the resulting array-level readout contrast is characterized by $\mathrm{CCR}_{\mathrm{cpl}} \approx 8.7\times10^{-5}$. This confirms that the differential-power readout is not a numerical artifact and persists in an N-body electromagnetic environment. Separately, the binding study shows that the prosthaphaeresis-plus-wrapping pipeline survives realistic propagation and recovers the target discrete operation under the stated aliasing plan. At the same time, the coupled regime is naturally susceptible to saturation because library--library near- and mutual-field interactions can dominate the local baseline power at each element.

Accordingly, the paper distinguishes the coupled (saturated) regime from an emitter-isolated regime and specifies an emitter-isolated RFC architecture in which inter-library coupling is suppressed so that $\Delta P$ is measured against an isolated self-power baseline. Mantle cloaking is presented as the primary intended physical realization of this anisotropic isolation requirement (e.g., strong suppression of in-plane scattering across the operating comb), but the paper does not claim mantle-cloak performance from cloak simulations. Instead, it validates the architectural premise of anisotropic isolation with a controlled surrogate experiment (Appendix~\ref{app:isolation_surrogate}) that suppresses the in-plane library--library path while preserving the off-plane query interaction, reducing the baseline scale by $32.67$~dB and increasing contrast by approximately $241\times$. This separation is deliberate: the paper establishes the physical computing principle and the architecture-level isolation requirement, while leaving cloak-specific metasurface synthesis and low-loss implementation as engineering work.

Overall, the contribution is a unified, physically grounded formulation of Resonant Field Computing: a regime in which core HDC-like operations arise within one coherent wave substrate and similarity serves as a measurable interface to that stack. The results separate what is algorithmically immediate (unitary embedding, physical realization of primitives, and the $\Delta P$ similarity mechanism) from what is engineering-limiting (coherence distribution, emitter isolation, and measurement sensitivity). The next steps are therefore clear: (i) experimental small-scale prototypes that directly measure $\Delta P$ under controlled coherence, (ii) systematic study of isolation mechanisms (including mantle cloaks as the target realization and surrogate isolators as a prototyping tool), and (iii) scaling studies that quantify CCR and readout time as functions of library size, isolation depth, and jitter budget.

\appendix
\section{Simulation Methodology}

This appendix describes the full 3D finite--difference time--domain (FDTD) methodology used to validate the Resonant Field Computing (RFC) correlation mechanism. It specifies the computational grid, antenna geometry, tone-comb construction, boundary conditions, solver settings, power-measurement procedure, normalization, and the definition of the Correlation Contrast Ratio (CCR). All reported simulation results in the main text are obtained using this configuration unless otherwise stated; in particular, the array is fully mutually coupled (no emitter isolation or mantle cloaks are modeled), so the reported CCR corresponds to the saturated comparison architecture.

\subsection{Computational Engine and Discretisation}

All full–wave simulations are performed with MEEP, an open–source FDTD solver that integrates Maxwell’s equations on a staggered Yee lattice in time and space \citep{oskooi2010meep,yee1966}. The electric field components $(E_x,E_y,E_z)$ and magnetic field components $(H_x,H_y,H_z)$ are placed on interleaved edges and faces of cubic cells, and advanced in time using second–order central differences.%

The simulations use MEEP’s dimensionless unit system with a normalized center frequency
\[
f_{\mathrm{cen}} = 1.0
\]
so that the free–space wavelength at the carrier is $\lambda_{\mathrm{cen}} = 1$ in simulation units.

The spatial resolution is fixed at
\[
\mathrm{resolution} = R = 10 \;\text{pixels per unit length},
\]
giving a uniform grid spacing
\[
\Delta x = \Delta y = \Delta z = \frac{1}{R} = 0.1.
\]
This choice ensures $\sim 10$ cells per wavelength at $f_{\mathrm{cen}}$, which is sufficient to control numerical dispersion while keeping the 3D problem tractable on commodity hardware.%

The time step $\Delta t$ is chosen automatically by MEEP according to the Courant–Friedrichs–Lewy stability condition for a 3D Yee grid, with a Courant factor $S \approx 0.5$:
\[
\Delta t = S\,\frac{\Delta x}{c\sqrt{3}},
\]
where $c$ is the speed of light in the normalized unit system.

\subsection{Simulation Domain and Boundary Conditions}

The RFC array is embedded in a finite rectangular domain with perfectly matched layer (PML) absorbing boundaries that emulate an unbounded free–space environment \citep{berenger1994,oskooi2010meep}. The computational cell has dimensions
\[
\texttt{cell\_size} = (s_{xy}, s_{xy}, s_z)
\]
with
\[
s_{xy} = 2\,(r_{\mathrm{array}} + l_{\mathrm{dipole}} + t_{\mathrm{PML}} + 2), \quad
s_z = w_{\mathrm{dipole}} + 2\,(t_{\mathrm{PML}} + 2),
\]
where
\[
r_{\mathrm{array}} = 5.0,\quad
l_{\mathrm{dipole}} = 0.5,\quad
w_{\mathrm{dipole}} = 0.02,\quad
t_{\mathrm{PML}} = 2.0.
\]
Numerically this gives $s_{xy} \approx 19.0$ and $s_z \approx 8.02$, so the grid is approximately $190 \times 190 \times 80$ cells at $R=10$.%

All six faces of the domain are surrounded by PML of thickness $t_{\mathrm{PML}} = 2.0$. The relatively thick PML (two wavelengths at $f_{\mathrm{cen}}$) is chosen because the correlation effect manifests as a very small normalized differential power, $\Delta \tilde P = O(10^{-4})$–$10^{-5}$; even weak boundary reflections could otherwise contaminate the measurement.

All simulations are run in full 3D (MEEP \texttt{dimensions=3}), enabling accurate modeling of polarization, radiation patterns, and mutual coupling between array emitters.

\subsection{Antenna Geometry and Array Layout}

Each emitter is modeled as a finite–length PEC dipole implemented with MEEP’s \texttt{mp.Block} primitive. The geometry of each antenna is
\[
\text{size} = (l_{\mathrm{dipole}}, \; w_{\mathrm{dipole}}, \; w_{\mathrm{dipole}})
= (0.5,\; 0.02,\; 0.02),
\]
and the material is \texttt{mp.metal}, which corresponds to a perfect electric conductor (PEC). All dipoles are oriented along the $x$–axis and centered at their respective positions.

The standard test configuration consists of one Query emitter $Q$ at the origin and a ring of 12 Library emitters $L_0, \dots, L_{11}$ arranged on a circle of radius $r_{\mathrm{array}} = 5.0$ in the $xy$–plane:
\[
\mathbf{r}_Q = (0, 0, 0),
\quad
\mathbf{r}_{L_i} = \big(r_{\mathrm{array}}\cos\theta_i,\; r_{\mathrm{array}}\sin\theta_i,\; 0\big),
\quad
\theta_i = \frac{2\pi i}{12}.
\]
All 13 dipoles share the same orientation (parallel $x$–directed dipoles), which maximizes radiative coupling while sampling multiple observation angles around the query source.

\subsection{Hypervector Dimensionality and Tone Comb}

Information is encoded in $N$–dimensional bipolar hypervectors
\[
\mathbf{x} \in \{-1,+1\}^N,
\]
with $N = 1000$ in the documented 3D calibration reported here. This dimensionality is large enough that two independent random bipolar hypervectors are quasi–orthogonal with high probability (mean cosine similarity $0$ and standard deviation $1/\sqrt{N} \approx 0.03$), which is essential for RFC’s similarity–based retrieval \citep{kanerva2009}.

The exact numerical values quoted for the 3D coupled-array validation are tied to this documented $N=1000$ configuration. Related runs with nearby parameter choices were also explored during development, but they are not used as hidden provenance for the specific headline numbers reported here.

Each dimension of the hypervector is mapped to one frequency bin in a dense tone comb used by the Unitary Wave Embedding (UWE). In the algebraic presentation of Section~3.1 this comb is written in centered form; in the representative 3D calibration reported here, the injected real waveform is synthesized from the positive-frequency half-spectrum. The fundamental spacing is chosen as
\[
f_0 = \frac{2 f_{\mathrm{cen}}}{N} = \frac{2}{1000} = 0.002,
\]
and the waveform uses the first $N/2$ positive frequencies to obtain a real–valued time-domain source signal. The effective occupied source bandwidth is therefore
\[
BW = \frac{N}{2} f_0 = 500 \times 0.002 = 1.0,
\]
yielding a multi-tone signal on the interval $[0, BW] = [0,1.0]$ in normalized units. This is a source-injection convention rather than a change to the underlying similarity algebra: moving the occupied comb to a different absolute spectral location changes carrier placement and calibration, while the encoded geometry is controlled by tone spacing, phase locking, and the chosen readout / wrapping convention.

\subsection{Signal Synthesis and Source Excitation}

Given a bipolar hypervector $\mathbf{x} = (x_0,\dots,x_{N-1})$, the representative 3D calibration workflow constructs the injected real waveform as follows \citep{oppenheim1999}:

\begin{enumerate}
  \item Compute the discrete Fourier transform (DFT):
  \[
  X_k = \mathrm{DFT}(\mathbf{x})_k, \quad k = 0,\dots,N-1.
  \]
  \item Extract the magnitudes and phases for the positive-frequency half-spectrum:
  \[
  A_k = |X_k|, \quad \phi_k = \arg(X_k), \quad k = 0,\dots,N/2.
  \]
  \item Synthesize a real–valued multi–tone waveform over a symbol period $T$:
  \[
  s(t) = \sum_{k=0}^{N/2} A_k \cos\big(2\pi k t/T + \phi_k\big),
  \]
  with
  \[
  T = \frac{N}{2 f_{\mathrm{cen}}} = \frac{1000}{2} = 500.
  \]
\end{enumerate}

This is the real-signal parametrization actually passed to \texttt{mp.CustomSource}. The centered-comb expression in Section~3.1 remains the algebraically transparent statement of the UWE; the simulation source uses its real, positive-frequency implementation form. In the implementation, this sum is evaluated on a fine time grid with step $\Delta t_{\mathrm{wave}}$ chosen such that several thousand samples span a single period $T$; the resulting array is wrapped modulo $T$ to produce a periodic source function $s(t)$ passed to \texttt{mp.CustomSource}.%

For each Library antenna $L_i$, a distinct hypervector $\mathbf{x}_i$ is drawn uniformly at random from $\{-1,+1\}^N$ and converted to a waveform $s_i(t)$ using the UWE mapping. The Query waveform $s_Q(t)$ is generated from a hypervector $\mathbf{x}_Q$ that either matches $\mathbf{x}_0$ (for the \emph{match} case) or is drawn independently (for the \emph{non–match} case).%

All sources are launched with the same start time $t=0$ inside the FDTD simulation, so that all tones are phase–locked across the array. Physically this corresponds to a shared clock or LO distribution network in hardware. Without this global phase coherence, the interference terms encoding similarity would average out and the RFC effect would vanish.

\paragraph*{Source Placement.}  
Each dipole is excited by a single \texttt{mp.CustomSource} current driven along its long axis (component \texttt{Ex}) and centered at the geometric center of the PEC block. The source extends a small finite length along the dipole (matching the metal block’s span in $x$), approximating a center–fed dipole.

\paragraph*{Simulation Duration.}  
Let $T_{\mathrm{wave}} = T$ denote the waveform period. Each simulation is run for
\[
t_{\mathrm{stop}} = 3 T_{\mathrm{wave}} = 1500
\]
plus an additional decay interval of one period, for a total duration of $t_{\mathrm{stop}} + T_{\mathrm{wave}} = 2000$ in simulation time units. This is sufficient for transient fields to decay and for the spectral flux monitors to converge to steady–state values.

\subsection{Power Measurement and Normalization}

The observable used by RFC is the radiated power associated with each Library antenna, obtained from the net Poynting flux through a closed box surrounding that antenna.

\paragraph*{Flux Box Definition.}  
For each Library dipole, a rectangular flux box of size
\[
\texttt{flux\_box\_size} = (l_{\mathrm{dipole}} + 0.2,\; w_{\mathrm{dipole}} + 0.2,\; w_{\mathrm{dipole}} + 0.2)
\]
is centered on the dipole. Six planar flux regions (one per face) are attached to this box. MEEP returns the spectral flux through each face, which is combined into a net power spectrum via
\[
P(\nu) = \big(F_{x+} - F_{x-}\big) + \big(F_{y+} - F_{y-}\big) + \big(F_{z+} - F_{z-}\big),
\]
where $F_{x+}$ and $F_{x-}$ denote the raw monitor outputs on the $+x$ and $-x$ faces, respectively, and similarly for $y$ and $z$. The plus/minus pairing converts the two opposing face measurements into a signed net outward flux along each axis before the three axes are summed.%

The spectra are recorded over $n_{\mathrm{freq}} = N/2$ frequency points spanning the positive-frequency source band, with center frequency and bandwidth
\[
f_{\mathrm{cen,flux}} = \frac{n_{\mathrm{freq}}}{2}\,f_0,
\quad
df_{\mathrm{flux}} = n_{\mathrm{freq}}\,f_0,
\]
which for the documented $N=1000$ calibration gives
\[
f_{\mathrm{cen,flux}} = 0.5,
\qquad
df_{\mathrm{flux}} = 1.0.
\]
Thus the flux monitors cover the interval $[0,1]$, matching the injected source band.

\paragraph*{Integrated Power.}  
For each Library emitter $L_j$, the total radiated power in a given run is obtained by numerically integrating its net spectrum:
\[
P_j = \sum_{m=1}^{n_{\mathrm{freq}}} P_j(\nu_m)\,\Delta \nu,
\quad
\Delta \nu = \frac{df_{\mathrm{flux}}}{n_{\mathrm{freq}}}.
\]

\paragraph*{Baseline and Query Runs.}  
To isolate the RFC correlation effect from bulk radiation, a two–run protocol is used:

\begin{itemize}
  \item \textbf{Baseline run.} Only the 12 Library emitters are driven (all $s_i(t)$ active, $s_Q(t)$ inactive). The integrated power at each Library antenna is recorded as $P_{j}^{\mathrm{(base)}}$.
  \item \textbf{Query run.} The same 12 Library emitters are driven, and the Query emitter is also driven with waveform $s_Q(t)$. The integrated power at each Library antenna is recorded as $P_{j}^{\mathrm{(query)}}$.
\end{itemize}

The differential power at Library element $L_j$ is then
\[
\Delta P_j = P_{j}^{\mathrm{(query)}} - P_{j}^{\mathrm{(base)}}.
\]

\paragraph*{Reference Power $P_{\mathrm{ref}}$.}  
Raw FDTD power values are in solver units and depend on resolution and source amplitude. To obtain a meaningful dimensionless quantity, a separate calibration simulation is performed with a single isolated dipole driven by the same waveform in an otherwise identical grid and PML configuration. The integrated power radiated by this isolated dipole is
\[
P_{\mathrm{ref}},
\]
and all differential powers are normalized as
\[
\Delta \tilde{P}_j = \frac{\Delta P_j}{P_{\mathrm{ref}}}.
\]
In this appendix and the associated tables, normalized quantities are written explicitly as $\Delta \tilde{P}_j$; $\Delta P_j$ is reserved for raw differential power in solver units.

\subsection{Correlation Contrast Ratio (CCR)}

The primary figure of merit extracted from the simulations is the Correlation Contrast Ratio (CCR), which measures how distinctly the system separates a matching hypervector from a non–matching one in terms of normalized differential power.

Let $\Delta \tilde{P}_{\mathrm{match}}$ denote the normalized differential power at the Library element whose hypervector matches the Query (i.e., $L_0$ with $\mathbf{x}_0 = \mathbf{x}_Q$), and let $\Delta \tilde{P}_{\mathrm{non}}$ denote the normalized differential power at a representative non–matching Library element (e.g., $L_1$). In the reported simulation, these values are approximately
\[
\Delta \tilde{P}_{\mathrm{match}} \approx +8.8 \times 10^{-5}, \quad
\Delta \tilde{P}_{\mathrm{non}} \approx -8.6 \times 10^{-5}.
\]

The Correlation Contrast Ratio is defined as
\[
\mathrm{CCR} =
\frac{\big|\Delta \tilde{P}_{\mathrm{match}} - \Delta \tilde{P}_{\mathrm{non}}\big|}
     {P_{0}^{\mathrm{(base)}}/P_{\mathrm{ref}} + P_{1}^{\mathrm{(base)}}/P_{\mathrm{ref}}},
\]
i.e., the absolute signal swing between match and non–match, normalized by the sum of their baseline powers. 

\noindent\textit{Remark.}
This expression is algebraically identical to Eq.~(9) in the main text; it is written 
here in terms of normalized powers 
$\Delta\tilde{P}_j = \Delta P_j / P_{\mathrm{ref}}$ and 
$P_j^{(\mathrm{base})}/P_{\mathrm{ref}}$ 
so the CCR values correspond directly to the FDTD results.

For the documented $N=1000$ calibration run used for the quoted coupled-array values, the corresponding normalized baseline sum for the representative match / non-match pair is approximately
\[
\frac{P_{0}^{\mathrm{(base)}}}{P_{\mathrm{ref}}} + \frac{P_{1}^{\mathrm{(base)}}}{P_{\mathrm{ref}}} \approx 2.0.
\]
Hence,
\[
\mathrm{CCR}_{\mathrm{cpl}} \approx \frac{\big|(+8.8\times 10^{-5}) - (-8.6\times 10^{-5})\big|}{2.0} \approx 8.7 \times 10^{-5}.
\]

For the configuration described above, this evaluates to
\[
\mathrm{CCR}_{\mathrm{cpl}} \approx 8.7 \times 10^{-5},
\]
demonstrating that the RFC mechanism produces a reproducible, bipolar (positive/negative) power redistribution in the physical electromagnetic system that is consistent with the similarity structure of the underlying hypervectors.

\subsection{Summary of Key Parameters}

For convenience, Table~\ref{tab:sim-params} summarizes the core parameters used in all RFC verification simulations.

\begin{table}[h]
\centering
\caption{Summary of simulation parameters used in RFC FDTD verification}
\label{tab:sim-params}
\begin{tabular}{ll}
\hline
\textbf{Category} & \textbf{Value / Description} \\
\hline
Domain dimensionality & 3D (MEEP \texttt{dimensions=3}) \\
Spatial resolution & $R = 10$ pixels / unit length \\
Cell size & $s_{xy} \approx 19.0$, $s_z \approx 8.02$ \\
PML thickness & $t_{\mathrm{PML}} = 2.0$ (all faces) \\
Dipole geometry & $0.5 \times 0.02 \times 0.02$ (PEC block, $x$–oriented) \\
Array layout & 12 Library + 1 Query, radius $r_{\mathrm{array}} = 5.0$ \\
Hypervector dimensionality & $N = 1000$ (bipolar $\{-1,+1\}^N$) \\
Fundamental spacing & $f_0 = 2 f_{\mathrm{cen}}/N = 0.002$ \\
Bandwidth & $BW = 1.0$ (using $N/2$ tones) \\
Flux monitor band & center $0.5$, width $1.0$ \\
Symbol period & $T_{\mathrm{wave}} = 500$ (normalized units) \\
Simulation duration & $t_{\mathrm{stop}} + T_{\mathrm{wave}} = 2000$ \\
Measurement & 6–face Poynting flux box per Library dipole \\
Normalization & $\Delta \tilde{P}_j = \Delta P_j / P_{\mathrm{ref}}$ \\
CCR (reported) & $\mathrm{CCR}_{\mathrm{cpl}} \approx 8.7 \times 10^{-5}$ \\
\hline
\end{tabular}
\end{table}

\section{Numerical Validation of RFC Binding}
\label{app:binding}

\subsection{Objective}

This appendix provides a numerical validation that the RFC binding operation, defined as the physical multiplication of Unitary Wave Embedded (UWE) waveforms followed by spectral wrapping via the aliasing kernel $\chi_{\mathrm{wrap}}$, reproduces the standard discrete HDC binding $z = x \odot y$ up to numerical precision.

\subsection{Setup: Discrete Model}

I first consider bipolar vectors $x, y \in \{-1,+1\}^N$ with
\begin{equation}
    z_{\text{target}} = x \odot y.
\end{equation}
Let $F$ denote the unitary DFT, and let $X = F x$, $Y = F y$ be their spectra. The UWE waveform for $x$ is constructed as
\begin{equation}
    s_x(t) = \sum_{k=0}^{N-1} X_k \, e^{j 2\pi f_k t},
\end{equation}
with a symmetric tone comb
\begin{equation}
    f_k = f_{\text{cen}} + \Bigl(k - \frac{N-1}{2}\Bigr)\, \Delta f,
    \qquad \Delta f = \frac{1}{T},
\end{equation}
and an analogous definition for $s_y(t)$. For the purely discrete experiment I choose moderate dimensionality ($N=32$) and parameters that ensure the tones remain resolvable over the window $T$; the specific values are not critical as long as the comb is coherent and bandlimited.

The discrete RFC binding pipeline is:

\begin{enumerate}
    \item Draw random bipolar vectors $x, y \in \{-1,+1\}^N$ and form
          $z_{\text{target}} = x \odot y$.
    \item Compute unitary DFTs $X = F x$ and $Y = F y$.
    \item Synthesize sampled UWE waveforms $s_x[n]$ and $s_y[n]$ on a grid of step $\Delta t$ over one symbol period $T$.
    \item Form the bound waveform $s_z[n] = s_x[n]\, s_y[n]$ (prosthaphaeresis).
    \item Compute the discrete-time Fourier transform of $s_z[n]$ to obtain a dense spectrum $S_z(f_i)$.
    \item Isolate the baseband (difference-frequency) region containing the linear convolution of $X$ and $Y$.
    \item Implement spectral wrapping by assigning each baseband bin at frequency $f_i$ to an $N$-bin index $k_i = \mathrm{round}(f_i / \Delta f)$ and accumulating its contribution into $Z_{k_i \bmod N}$:
    \begin{equation}
      Z_k^{\mathrm{RFC}}
      \;=\;
      \sum_{i:\, \mathrm{round}(f_i/\Delta f) \equiv k \ (\mathrm{mod}\ N)}
      S_z(f_i),
    \end{equation}
    which is the discrete aliasing kernel $\chi_{\mathrm{wrap}}$.
    \item Treat $Z^{\mathrm{RFC}}$ as the binding spectrum and apply an inverse DFT to recover $z_{\mathrm{RFC}} \in \mathbb{C}^N$.
    \item Compare $z_{\mathrm{RFC}}$ to $z_{\text{target}}$ using cosine similarity and sign agreement.
\end{enumerate}

A reference Python/NumPy implementation used for this experiment is given in Listing~\ref{lst:rfc_binding-discrete}. All operations are performed in double-precision floating point.

\begin{lstlisting}[language=Python, caption={Discrete RFC binding pipeline with explicit spectral wrapping.}, label={lst:rfc_binding-discrete}]
import numpy as np

# 1. Generate bipolar HDC vectors and target
N = 32
rng = np.random.default_rng(42)
x = rng.choice([-1, 1], size=N)
y = rng.choice([-1, 1], size=N)
z_target = x * y

# 2. Unitary DFT
Fx = np.fft.fft(x, norm="ortho")
Fy = np.fft.fft(y, norm="ortho")

# 3. UWE waveform synthesis
f_cen, delta_f = 2.4e9, 1.0e6
T_window = 1.0 / delta_f
fs = 12.0e9
num_samples = int(T_window * fs)
t = np.arange(num_samples) / fs

def uwe_waveform(spectrum, f_cen, delta_f, t_vec):
    N = len(spectrum)
    sig = np.zeros_like(t_vec, dtype=np.complex128)
    for k in range(N):
        f_k = f_cen + (k - (N - 1) / 2) * delta_f
        sig += spectrum[k] * np.exp(1j * 2 * np.pi * f_k * t_vec)
    return sig.real

s_x = uwe_waveform(Fx, f_cen, delta_f, t)
s_y = uwe_waveform(Fy, f_cen, delta_f, t)

# 4. Physical binding (time-domain multiplication)
s_z = s_x * s_y

# 5. Spectrum and baseband selection
Fz_full = np.fft.fft(s_z)
freqs_full = np.fft.fftfreq(num_samples, d=1/fs)

max_baseband = N * delta_f
mask = np.abs(freqs_full) < max_baseband
Fz_bb = Fz_full[mask]
freqs_bb = freqs_full[mask]

# 6. Spectral wrapping (aliasing kernel)
target_bins = np.round(freqs_bb / delta_f).astype(int)
Fz_wrapped = np.zeros(N, dtype=np.complex128)
for val, idx in zip(Fz_bb, target_bins):
    Fz_wrapped[idx % N] += val

# 7. Invert to recover vector
z_rfc = np.fft.ifft(Fz_wrapped, norm="ortho").real

# 8. Cosine similarity
cos_sim = np.dot(z_target, z_rfc) / (
    np.linalg.norm(z_target) * np.linalg.norm(z_rfc)
)
print("Cosine similarity:", cos_sim)
\end{lstlisting}

For representative runs, the recovered vector $z_{\mathrm{RFC}}$ is numerically indistinguishable from $z_{\text{target}}$ up to floating-point precision (cosine similarity $\approx 1.0$ and identical sign pattern).

\subsection{Setup: FDTD-Based Binding with Frequency Planning}

To demonstrate that the same aliasing kernel can be realized in a physically propagated field, I perform a second experiment using a finite–difference time–domain (FDTD) simulation.

I choose $N_{\mathrm{dim}} = 128$ and a tone spacing $\Delta f = 0.01$ (normalized units). The center frequency is set to
\begin{equation}
  f_{\mathrm{cen}} = 2.5,
\end{equation}
so that the optimized comb occupies
\begin{equation}
  f_k = f_{\mathrm{cen}} + \Bigl(k - \frac{N_{\mathrm{dim}} - 1}{2}\Bigr)\Delta f,
\end{equation}
yielding a signal band
\[
  f_{\mathrm{signal}} \in [1.86, 3.13],
\]
a mixer \emph{difference} band (data) approximately
\[
  f_{\mathrm{diff}} \in [0.00, 1.27],
\]
and a mixer \emph{sum} band (interference)
\[
  f_{\mathrm{sum}} \in [3.73, 6.27].
\]
This frequency planning ensures that the difference band and sum band are spectrally separated so a simple low-pass filter (e.g.\ cutoff near $f=2.0$) can isolate the data-bearing difference term.

Two random bipolar hypervectors $x,y \in \{-1,+1\}^{N_{\mathrm{dim}}}$ are embedded into spectra $X,Y$ and converted to continuous sources via UWE. Each source is injected into a 2D Meep simulation (PEC-free homogeneous space with PML boundaries, resolution 50, cell size $20 \times 10$). The fields are recorded at a receiver point a distance $d=10$ units away from the source; the resulting time series are denoted $s_x(t)$ and $s_y(t)$.

The RFC binding pipeline implemented in this experiment is:

\begin{enumerate}
    \item Generate $x,y$ and $z_{\mathrm{target}} = x \odot y$.
    \item Build UWE sources from $x$ and $y$ on the frequency-planned comb.
    \item Run two FDTD simulations (one per source), recording the received fields $s_x(t)$ and $s_y(t)$ at the same spatial point.
    \item For a range of delays around the propagation time $d$, sample synchronized windows of length $T = 1/\Delta f$:
    \[
      s_x^{(d)}[n] = s_x(t_0^{(d)} + n \Delta t),
      \quad
      s_y^{(d)}[n] = s_y(t_0^{(d)} + n \Delta t).
    \]
    \item For each delay candidate, form the bound waveform $s_z^{(d)}[n] = s_x^{(d)}[n]\,s_y^{(d)}[n]$.
    \item Compute the FFT $F_{\mathrm{mix}}^{(d)}[i]$ and corresponding frequency grid $f_i$.
    \item Apply a low-pass filter to keep only $|f_i| \leq f_{\mathrm{cut}}$ with $f_{\mathrm{cut}} = 2.0$, rejecting the sum band.
    \item Implement the aliasing kernel $\chi_{\mathrm{wrap}}$ via modulo-$N_{\mathrm{dim}}$ binning:
    \[
      k_i = \mathrm{round}(f_i / \Delta f),
      \qquad
      Z_{\mathrm{accum}}[k_i \bmod N_{\mathrm{dim}}]
      \mathrel{+}= F_{\mathrm{mix}}^{(d)}[i].
    \]
    \item Apply an inverse DFT to obtain $z_{\mathrm{RFC}}^{(d)} = F^{-1}\{Z_{\mathrm{accum}}\}$.
    \item Select the delay $d^\star$ that maximizes the cosine similarity
    \[
      \cos\theta^{(d)} = \frac{\langle z_{\mathrm{RFC}}^{(d)}, z_{\mathrm{target}}\rangle}{\|z_{\mathrm{RFC}}^{(d)}\|\,\|z_{\mathrm{target}}\|}.
    \]
\end{enumerate}

The core of the implementation matches the discrete aliasing kernel in \eqref{eq:chi-wrap-continuous}–\eqref{eq:chi-wrap-ideal}, but now applied to spectra obtained from physically propagated fields.

\subsection{Results}

For typical runs with the frequency-planned comb described above, the best-delay reconstruction satisfies
\[
\cos\theta^{(d^\star)} \approx 0.9990,
\]
and the recovered vector
\[
z_{\mathrm{bin}} = \mathrm{sign}\bigl(z_{\mathrm{RFC}}^{(d^\star)}\bigr)
\]
matches the target bound vector exactly on all components, i.e.\ the bit accuracy is
\[
\mathrm{Acc}_{\mathrm{bits}} = \frac{1}{N_{\mathrm{dim}}}\sum_{i=1}^{N_{\mathrm{dim}}} \mathbb{1}\{z_{\mathrm{bin},i} = z_{\mathrm{target},i}\} = 100\%.
\]

These results demonstrate that:

\begin{itemize}
  \item The combination of UWE, physical propagation, prosthaphaeresis (time-domain multiplication), and spectral wrapping via the aliasing kernel $\chi_{\mathrm{wrap}}$ reproduces the discrete HDC binding operation in the embedded space.
  \item Frequency planning that separates the sum and difference bands is sufficient to allow a simple low-pass filter plus modulo-$N$ binning to implement the aliasing kernel.
  \item The binding pipeline remains numerically stable under realistic field propagation, with cosine similarity $\approx 0.999$ and perfect sign recovery in the tested configuration.
\end{itemize}

\subsection{Summary}

Taken together, the purely discrete and FDTD-based experiments confirm that the RFC binding pipeline,
\[
x,y \;\mapsto\; s_x(t), s_y(t) \;\mapsto\; s_x(t)s_y(t) \;\mapsto\; Z_{\mathrm{circ}}(\omega) \;\mapsto\; z_{\mathrm{RFC}},
\]
implements the same algebraic operation as discrete HDC binding $z = x \odot y$ when viewed in the embedded space. The aliasing kernel $\chi_{\mathrm{wrap}}$, derived in the main text via Poisson summation and realized numerically as modulo-$N$ frequency re-binning, closes the gap between linear physical mixing and circular convolution, and the FDTD experiment shows that this mechanism survives realistic wave propagation.

\section{Numerical Validation of RFC Permutation}
\label{app:permutation}

\subsection{Objective}
This appendix validates the claim that the RFC permutation primitive, implemented physically as a time delay, reproduces the standard discrete HDC cyclic shift operation. Specifically, I demonstrate that the Unitary Wave Embedding (UWE) of a cyclically shifted vector is equivalent to the time-delayed UWE waveform of the original vector.

\subsection{Theory}
In discrete HDC, permutation is often defined as a cyclic shift of the vector coordinates. Let $\Pi^k$ denote a cyclic shift by $k$ positions. For a vector $x \in \{-1, +1\}^N$:
\begin{equation}
    (\Pi^k x)_n = x_{(n-k) \pmod N}.
\end{equation}
Under the Unitary Wave Embedding defined in Section 3.1, the time-domain waveform is a sum of Fourier components. By the Fourier Shift Theorem, a shift in the discrete sequence corresponds to a linear phase shift in the frequency domain. Physically, this manifests as a time delay $\tau$. The relationship between the discrete shift index $k$ and the physical delay $\tau$ is given by:
\begin{equation}
    \tau = k \cdot \frac{T}{N},
\end{equation}
where $T$ is the symbol period and $N$ is the hypervector dimensionality.

\subsection{Procedure}
The numerical validation compares two methods of generating the permuted waveform:
\begin{enumerate}
    \item \textbf{Method A (Discrete-First):} Apply the cyclic shift $k$ to the discrete hypervector $x$ to get $x'$, then generate the UWE waveform $s_{A}(t)$ from $x'$.
    \item \textbf{Method B (Continuous-First):} Generate the UWE waveform $s(t)$ from the original $x$, then apply a time delay $\tau$ to generate $s_{B}(t) = s(t - \tau)$. In the sampled implementation below, this is realized as a positive circular sample shift.
\end{enumerate}

The simulation parameters match the standard configuration: $N=1024$, cyclic shift $k=50$, and waveform duration $T=1.0$ (normalized units). The equivalence is quantified by the Mean Squared Error (MSE) between $s_{A}(t)$ and $s_{B}(t)$. Additionally, I verify the \textbf{quasi-orthogonality} property by measuring the cosine similarity between the original and permuted representations in both domains.

A reference implementation of this experiment is shown in Listing~\ref{lst:rfc_permutation}.

\begin{lstlisting}[language=Python, caption={Python implementation of the RFC Permutation validation.}, label={lst:rfc_permutation}]
import numpy as np
from scipy.fft import fft

# Parameters
N = 1024
shift_k = 50
T = 1.0
dt = 1 / (2 * N) # Nyquist sampling
t_axis = np.arange(0, T, dt)

# 1. Generate Source
x = np.random.choice([-1, 1], size=N)

# 2. UWE Synthesis Function
def uwe_synthesis(vec, t_vals):
    X_k = fft(vec)
    # Using positive frequencies for synthesis (simplified)
    freqs = np.fft.fftfreq(N, d=1.0/N) * (1.0/T)
    s_t = np.zeros_like(t_vals)
    for k in range(N // 2 + 1):
        amp = np.abs(X_k[k])
        phase = np.angle(X_k[k])
        scale = 1/N if k==0 else 2/N
        s_t += scale * amp * np.cos(2*np.pi*freqs[k]*t_vals + phase)
    return s_t

# 3. Method A: Permute Discrete -> Encode
x_permuted = np.roll(x, shift_k)
s_method_A = uwe_synthesis(x_permuted, t_axis)

# 4. Method B: Encode -> Time Shift
s_original = uwe_synthesis(x, t_axis)
tau = shift_k * (T / N)
# Apply circular time delay to waveform samples
shift_samples = int(tau / dt)
s_method_B = np.roll(s_original, shift_samples)

# 5. Metrics
mse = np.mean((s_method_A - s_method_B)**2)
sim_discrete = np.dot(x, x_permuted) / N
# Normalized cross-correlation for waveforms
norm_factor = np.sqrt(np.sum(s_original**2) * np.sum(s_method_A**2))
sim_continuous = np.correlate(s_original, s_method_A)[0] / norm_factor
\end{lstlisting}

\subsection{Results}
The experiment confirms the duality between discrete permutation and continuous time delay.

\begin{table}[h]
    \centering
    \caption{Results of the RFC Permutation validation.}
    \begin{tabular}{lcl}
        \toprule
        Metric & Value & Interpretation \\
        \midrule
        \textbf{Equivalence (MSE)} & $1.81 \times 10^{-4}$ & Methods A and B are numerically very close at the reported discretization. \\
        \textbf{Discrete Orthog.} & $-0.0117$ & Permuted vector is quasi-orthogonal to original. \\
        \textbf{Waveform Orthog.} & $-0.0127$ & Physical waveforms maintain orthogonality. \\
        \bottomrule
    \end{tabular}
\end{table}

The small Mean Squared Error supports the expected equivalence between discrete cyclic shifting and continuous time shifting at the reported discretization. Furthermore, the low correlation values confirm that the permutation successfully creates a distinct, quasi-orthogonal vector in the continuous domain, satisfying the requirements for role-filler binding in Vector Symbolic Architectures \citep{gayler2004}.

\section{Phase-Jitter Robustness Experiment (Binding)}
\label{app:jitter}
This appendix supports the phase-jitter tolerance discussion in Section~\ref{sec:jitter-binding}.

\subsection{Jitter Injection Model}
To test sensitivity to coherence loss, I perturb the UWE spectral phases prior to waveform synthesis. For an embedded spectrum \(X_k\), I form
\[
\widetilde{X}_k = X_k e^{j\delta\phi_k},
\qquad \delta\phi_k \sim \mathcal{N}(0,\sigma_\phi^2),
\]
independently across tones \(k\). The same perturbation model is applied to both operands \(x\) and \(y\) (with independent draws unless otherwise stated). This corresponds to relative per-tone phase error (e.g., uncompensated group-delay ripple, calibration error, or multi-path phase distortion across the comb), and is therefore a conservative impairment relative to purely common-mode LO phase noise.

\subsection{Protocol and Metrics}
For each \(\sigma_\phi\), I run the FDTD binding pipeline (UWE synthesis \(\to\) propagation \(\to\) time-domain multiplication \(\to\) low-pass isolation of the difference band \(\to\) modulo-\(N\) spectral wrapping) and recover \(\hat{z}\). I report:
(i) cosine similarity \(\cos(z_{\mathrm{target}},\hat{z})\) and
(ii) sign accuracy \(\frac{1}{N}\sum_i \mathbb{1}\{\mathrm{sign}(\hat{z}_i)=z_{\mathrm{target},i}\}\).

\subsection{Representative Results}
A representative sweep produced:
\[
\sigma_\phi=\{0,0.1,0.2,0.5,1.0\}\ \mathrm{rad}
\Rightarrow
\cos(z_{\mathrm{target}},\hat{z})=\{0.9987,0.9942,0.9782,0.8539,0.4378\},
\]
with corresponding sign accuracies \(\{100,100,100,96.88,67.19\}\%\).
These results demonstrate a robust regime (\(\sigma_\phi\le 0.2\) rad) and graceful degradation beyond, consistent with standard HDC noise tolerance \citep{kanerva2009,kleyko2022survey}.

\section{Isolation Surrogate Simulation (2D Anisotropic Baffle)}
\label{app:isolation_surrogate}

\subsection{Objective}
This appendix reports a controlled surrogate experiment that enforces anisotropic emitter isolation in order to validate the
architectural claim that suppressing library--library coupling lifts saturation of the $\Delta P$ readout. The isolator used here
is a lossy baffle and is \emph{not} a scattering-cancellation mantle cloak.

\subsection{Geometry and sources}
I simulate three radiators in 2D: a target library emitter $L_1$ at the origin, a neighboring library emitter $L_2$ placed along
the array plane ($x$ axis), and a query source $Q$ placed off-plane ($y$ axis). An anisotropic isolator is implemented as a lossy
baffle placed only between $L_1$ and $L_2$, suppressing the in-plane L--L path while leaving the off-plane Q--$L_1$ path unobstructed.
All three sources are driven by UWE multi-tone waveforms derived from bipolar hypervectors (here using $N_{\mathrm{dim}}=128$ for runtime).

\subsection{Measurement protocol}
At $L_1$ I measure the net outward Poynting flux through a closed box surrounding $L_1$, constructed from four face flux monitors
(2D analog of the 3D flux-box methodology). For each configuration I run:
(i) a baseline run with library emitters active and query off, giving $P^{(\mathrm{base})}$, and
(ii) a query run with query on, giving $P^{(\mathrm{query})}$.
I define the differential power proxy as
\[
\Delta P = P^{(\mathrm{query})} - P^{(\mathrm{base})}.
\]
I evaluate both a match query ($Q$ generated from the same hypervector as $L_1$) and a mismatch query ($Q$ generated from an independent random hypervector).

\subsection{CCR-like contrast metric}
For this single monitored emitter surrogate, I report the contrast ratio
\[
\mathrm{CCR}_{\mathrm{sur}} =
\frac{\big|\Delta P_{\mathrm{match}}-\Delta P_{\mathrm{mis}}\big|}{2\,|P^{(\mathrm{base})}|},
\]
where the absolute value avoids sign ambiguities of net flux under interference.

\subsection{Numerical results}
In the unisolated (no baffle) surrogate configuration I obtained:
\[
P^{(\mathrm{base})} = -2.342078\times 10^{-3},\quad
\Delta P_{\mathrm{match}} = -2.061490\times 10^{-1},\quad
\Delta P_{\mathrm{mis}} = -7.022430\times 10^{-2},
\]
which yields $\mathrm{CCR}_{\mathrm{sur}} \approx 2.90\times 10^{1}$.
With the anisotropic baffle enabled I obtained:
\[
P^{(\mathrm{base})} = -1.266211\times 10^{-6},\quad
\Delta P_{\mathrm{match}} = -3.614158\times 10^{-2},\quad
\Delta P_{\mathrm{mis}} = -1.843098\times 10^{-2},
\]
which yields $\mathrm{CCR}_{\mathrm{sur}} \approx 6.99\times 10^{3}$.
Thus, the baseline scale decreases by $32.67$~dB and the contrast ratio increases by approximately $241\times$.

\subsection{Interpretation}
This surrogate demonstrates the architecture-level point: suppressing coupling-dominated baseline power can substantially lift
$\Delta P$ contrast, moving the system from a saturated N-body regime toward many parallel query--library interactions. Because the
isolator is lossy, this result does not quantify metasurface mantle-cloak scattering-cancellation performance; it validates the
role that mantle cloaking is intended to realize with lower loss and better scalability.

\section*{Use of Generative AI Tools}
Generative AI language tools were used extensively to assist with drafting, restructuring, and editing the manuscript under the author's direction. All technical claims, derivations, citations, simulations, and conclusions were reviewed by the author, who assumes full responsibility for the content.

\bibliographystyle{unsrtnat}
\bibliography{references}
\end{document}